\documentclass[prb, twocolumn,superscriptaddress,preprintnumbers,a4paper,amsmath,amssymb,showpacs,floatfix]{revtex4}
\usepackage{graphicx}
\usepackage{epstopdf}
\usepackage{amsmath,amssymb,graphicx,bm,epsfig}

\newcommand{\vk}{{\mathbf{k}}}

\begin{document}

\title{Kink during the formation of the Kondo resonance band in the heavy fermion system}

\author{Hong Chul Choi}
\affiliation{Department of Physics,
Pohang University of Science and Technology, Pohang 790-784, Korea}
\affiliation{Department of Chemistry,
Pohang University of Science and Technology, Pohang 790-784, Korea}
\author{K. Haule}
\affiliation{Department of Physics,
Rutgers University, Piscataway, NJ 08854, USA}	
\author{G. Kotliar}
\affiliation{Department of Physics,
Rutgers University, Piscataway, NJ 08854, USA}
\author{B. I. Min}
\email[]{bimin@postech.ac.kr}
\affiliation{Department of Physics,
Pohang University of Science and Technology, Pohang 790-784, Korea}
\author{J. H. Shim}
\email[]{jhshim@postech.ac.kr}
\affiliation{Department of Physics,
Pohang University of Science and Technology, Pohang 790-784, Korea}
\affiliation{Department of Chemistry,
Pohang University of Science and Technology, Pohang 790-784, Korea}
\affiliation{Division of Advanced Nuclear Engineering,
Pohang University of Science and Technology, Pohang 790-784, Korea}
\date{\today}

\begin{abstract}
We have shown that the kink behavior in the spectral function of heavy fermion can appear 
during the formation of the Kondo resonance (KR) band and the hybridization gap (HG).
We have investigated the heavy fermion compound CeCoGe$_2$, using a combined approach of
the density functional theory (DFT) and the dynamical mean field theory (DMFT).
Low temperature ($T$) spectral functions show
dispersive KR states, similarly to the recent experimental observation.
During the evolution from the $nonf$ conduction band state at high $T$ to the dispersive KR band state at low $T$,
which have topologically  different band shapes,
we have found the existence of kinks in the $nonf$ spectral function near $E_F$.
The observation of kink is clearly in correspondence with 
the multiple temperature scales of the formation of the KR band.
\end{abstract}

\pacs{71.20.-b, 71.27.+a, 75.30.Mb}

\maketitle

The Hamiltonians for the heavy fermion compounds, 
such as the Kondo lattice model and the periodic Anderson lattice model,\cite{Coleman}
are represented  
in terms of the kinetic energies of conduction electrons,
the correlation energies among localized electrons,
and the hybridization between them.
In Ce-based compounds, the localized electrons and the conduction electrons correspond to
Ce $4f$ electrons, and other $spd$ electrons.
The $nonf$ electrons are mostly metallic (delocalized) and well described by the density functional theory (DFT).
The localization of $4f$ electrons in the ground state is determined 
by the competition between the Ruderman-Kittel-Kasuya-Yosida (RKKY) interaction and Kondo effect.\cite{Coleman}
The RKKY interaction is the conduction electron-mediated exchange energy between $4f$ electrons.
The Kondo effect induces the spin singlet state of the localized and delocaled electrons.
The hybridization strength between $4f$ and $nonf$ electrons  strongly affects both the RKKY interaction and the Kondo effect.
At the weak hybridization, the RKKY interaction drives the localized 4f electrons to be magnetic.
The Kondo effect at the strong hybridization produces the Fermi liquid states 
with heavy mass.

Depending on the temperature, the localization of 4f electrons are changed in heavy fermion compounds, 
at high temperature ($T$), the 4f states are well localized to have negligible contribution to 
the Fermi level ($E_F$), and so only $nonf$ dispersive conduction bands are
observed near $E_F$ with effectively small hybridization.
At low $T$, however the Kondo effect overwhelms the RKKY interaction,
and, the $4f$ states appear near $E_F$ and 
to be hybridized with other conduction electrons.
Due to the flatness of the hybridized bands, 
the effective mass of quasiparticles becomes tens to 
hundreds times larger than the bare electron mass.
This quasiparticle band near $E_F$ can be identified 
as the dispersive Kondo resonance (KR) peak\cite{Allen} in the angle-resolved photoemission spectrum (ARPES).\cite{Im2008} 
The upper and lower hybridized bands produce 
The hybridization gap\cite{Coleman} (HG), 
which can be recognized below the coherent temperature ($T^{*}$),
when the heavy quasiparticle bands begin to emerge.

The HG feature was directly observed
by the optical conductivity,\cite{Marabelli,Dordevic,CeIn3,Kwon,UPt3,URu2Si2}
and described theoretically by
the periodic Anderson model.\cite{Coleman}
The optical conductivities show the mid-infrared peaks in the energy range of several-tens meV at low $T$.
These peaks are suppressed with increasing $T$ and 
disappear well above $T^{*}$
to produce the incoherent Drude peak at zero frequency.
The magnitude of the HG ($\Delta_{HG}$) measures
the hybridization strength between $f$ and other $nonf$ conduction states.
Recently, the dynamical mean field theory (DMFT) approach revealed 
the formation of the HG in CeIrIn$_5$ by calculating $T$-dependent spectral function $A(k,\omega)$.\cite{shim07}
The momentum-dependent $\Delta_{HG}$ at low $T$ was in good agreement with the measured spectrum 
and provided the information of the orbital-dependent hybridization strength.

\begin{figure}[t]
\includegraphics[scale=0.2,angle=0]{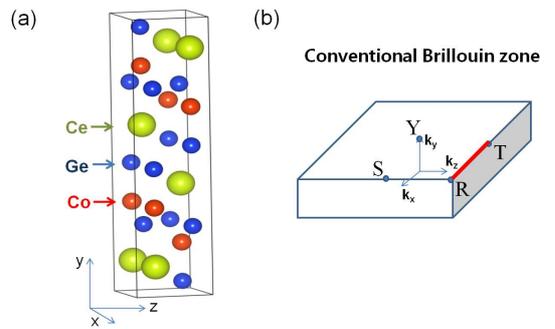}
\caption{(color online)
Crystal structure and its Brillouin zone
(a) The crystal structure of CeCoGe$_2$.\cite{vesta}
(b) The conventional Brillouin zone. Red line represents the path used in Fig.~\ref{fig2}.
}
\label{fig0}
\end{figure}

In this study, we have theoretically described the
formation of the dispersive KR band and the HG
in the heavy fermion compound CeCoGe$_2$.\cite{Mun1}
It has the orthorhombic base-centered structure, shown in Fig.~\ref{fig0}.
Though the nearest neighbours of Ce atoms are Ge atoms,
the $sp$ states in Ge ions are almost empty.
The open-core calculation ($4f$ states set to be inside the core),
which is equivalent to the electronic structure at high $T$,
demonstrated that Co $3d$ states have the most contribution to $E_F$.
So, Co $3d$ states can be regarded as main character of $nonf$ states.
The KR states are composed of dispersive bands
arising from the hybridization of $nonf$ and renormalized $f$ bands, and
show quantitative agreement with the recent ARPES measurement.\cite{Im2008}
We have shown that the formation of the KR band,
which brings about a topological change of the band structures, 
should be accompanied by the kinks in the spectral function.
We propose that
the kink could appear in the $nonf$ spectral function ($A^{nonf} (\vk,\omega)$)
prior to the formation of the KR band around $T^{*}$.

The kink, the abrupt change in the band dispersion, is usually observed in the ARPES measurements
of high $T_C$ superconductors.\cite{Lanzara,Gweon,Schachinger,Dahm,Hwang}
The discontinuous quasiparticle band 
is induced by various collective excitations, such as
phonon\cite{Lanzara,Gweon,Schachinger} and spin-fluctuation.\cite{Dahm}
When the strong electron-phonon coupling disturbs the velocity and the scattering rate of electrons,
ARPES spectrum\cite{Hengsberger,Valla,Rotenberg} near the phonon energy shows
the abrupt change in the slope of the energy-momentum dispersion.
The ARPES experiment on USb$_2$ also revealed the kink feature in the dispersion of $f$ bands, 
which was explained by the combination of quasiparticle bands and many-body correction 
of the electron-spin-fluctuation coupling.\cite{Durakiewicz}
On the other hand, it was reported
that the pure electronic correlation in the Hubbard model can produce such a kink near $E_F$,\cite{Byczuk}
which suggests that 
the one-particle picture should be strongly renormalized below 
a certain energy $\omega^{*}$ to have the kink feature in momentum space.
This analogy was applied to the periodic Anderson model, showing the existence of the kink.\cite{Kainz}
Recently, the energy scale of the kink was proved to be smaller than the width of the central peak.\cite{Held}
Other subsequent studies, however, showed that the kink in the Hubbard model arises from
the internal spin-fluctuation mode.\cite{Macridin,Chakraborty,Raas}
In the present study, instead of kinks in correlated bands,
we show that kinks can be phenomenologically observed in noncorrelated bands 
due to the formation of correlated $f$ states in the heavy fermion system.


We have used the DFT+ DMFT method,\cite{dmft}
implemented in the linearized Muffin-Tin orbital band method.\cite{Savrasov}
We have considered the experimental crystal structure and the Brillouin zone shown in Fig. 1.
The DMFT calculation considers only the local self-energy of $4f$ orbital,
and other orbitals are considered in the DFT part.
The local correlation effect is contained in the self-energy $\Sigma(\omega)$, which can be calculated from
the corresponding impurity problem.
To solve the impurity problem, we used a vertex corrected one-crossing approximation (OCA),\cite{dmft}
which is a self-consistent diagrammatic method perturbed in the atomic limit.\cite{Cowan}
We used the same interaction parameters U=5.0 and J=0.8 eV, as in previous works on CeIrIn5.\cite{shim07,hcc}
It has been checked that the OCA describes well the $T$-dependent spectral function of heavy fermion compound.\cite{shim07,hcc}
We neglected the crystalline electric field (CEF) effect on the local correlation energy 
because CeCoGe$_2$ has been confirmed as $j= 5/2$ heavy fermion.\cite{Mun1}

\begin{figure}[t]
\begin{center}$
\begin{array}{cc}
\includegraphics[scale=0.45,angle=0]{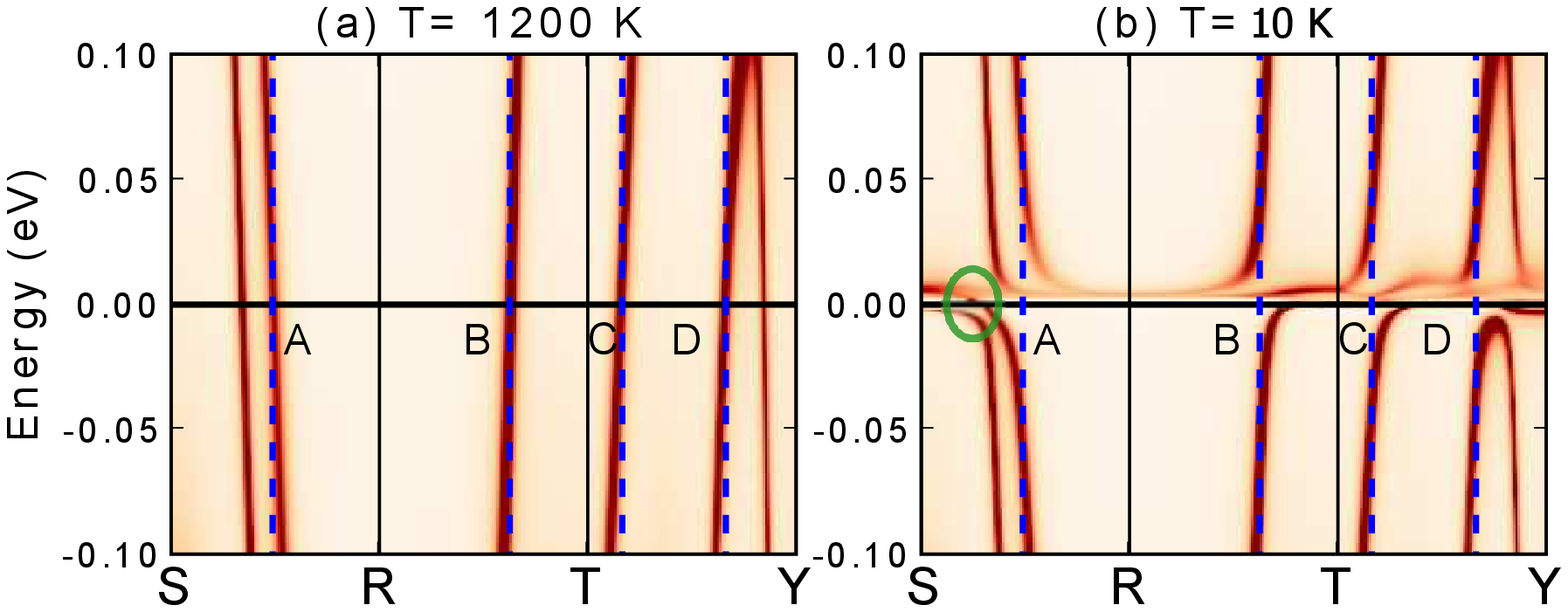}\\
\includegraphics[scale=0.30,angle=0]{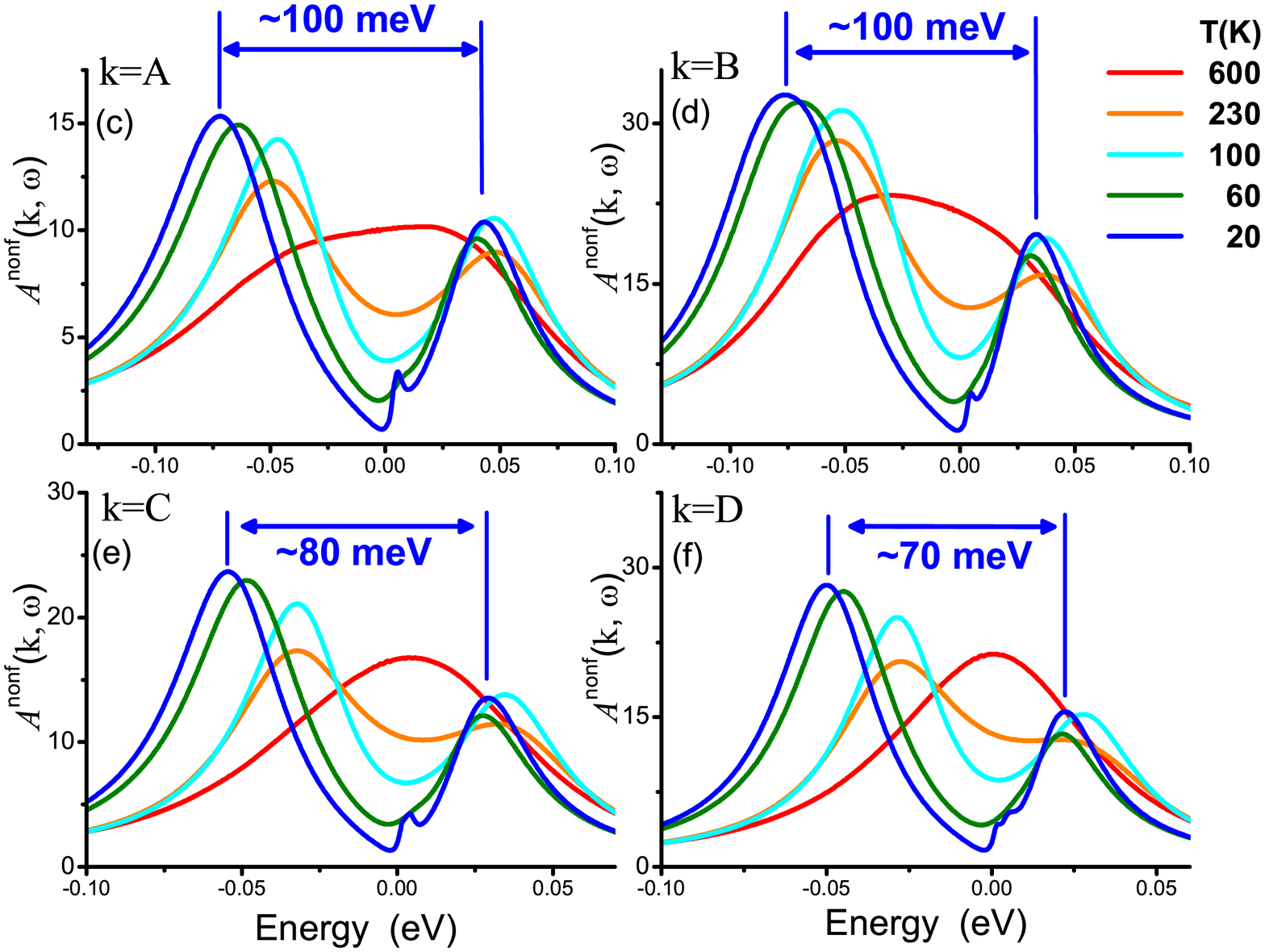}
\end{array}$
\end{center}
\caption{(color online)
Theoretical demonstration of the HG feature from $T$-dependent $A^{nonf} (\vk,\omega)$.
$A^{nonf} (\vk,\omega)$'s are provided along $S$-$R$-$T$-$Y$ 
(a) at $T=1200$ K, and (b) at $T=10$ K.
$S$, $R$, $T$, and $Y$ are $\vk=(\pi/2,\pi/2,0)$, $(\pi/2,\pi/2,\pi/2)$, $(0,\pi/2,\pi/2)$,
and $(0,\pi/2,0)$, respectively.
The green circle in (b) is for the emphasis of the $T$-dependent change.
Blue dotted lines marked by A, B, C, and D represent $\vk$-points,
where the $T$-dependent $A^{nonf} (\vk,\omega)$ calculations in (c)-(f) are done.
Theoretically predicted $\Delta_{HG}$ is given at each $\vk$-point.
}
\label{fig4}
\end{figure}

Figures~\ref{fig4}(a) and ~\ref{fig4}(b) show $A^{nonf} (\vk,\omega)$'s
along $S$-$R$-$T$-$Y$ at $T=1200$ and 10 K, 
which demonstrates the formation of the KR bands  near $E_F$.
The dispersive $spd$ bands at high $T$ (1200 K) look almost vertical
due to the narrow energy range.
At low $T$, as a result of the hybridization,
new coherent quasiparticle bands are formed to give the different band geometry near $E_F$.
For example, there are two separate bands crossing $E_F$ between $S$ and $R$ at high $T$,
while, at low $T$, additional $j=5/2$ bands with the bandwidth of $\sim$10 meV are introduced to produce
degenerated three composite quasiparticle bands.
As a result, the lower part of two bands observed at high $T$ are warped due to the hybridization,
as indicated by green circle in Fig.~\ref{fig4}(b).
One band closer to $S$ is pushed down below $E_F$, but another band still intersects $E_F$.
It is interesting that the formation of the parabolic-like hybridized band originated from high $T$ separated bands  below $E_F$ around $k=$ D at low $T$.
All those features result in the change of geometry in the FS, as will be shown in Figs.~\ref{fig5}(e) and \ref{fig5}(f).

$T$-dependent formation of the HG has been examined at the chosen $\vk$-points in Figs.~\ref{fig4}(c)-(f).
At high $T$, $A^{nonf}(\vk,\omega)$ shows a general quasiparticle spectral feature
with a single Gaussian function at each $k$-point.
With lowering $T$, the spectral weights of $A^{nonf} (\vk,\omega)$ near $E_F$
begin to be transferred to upper and lower peaks separated
by $\Delta_{HG}$ to form a gap structure.
At low $T$, the clear gap can be observed 
at each chosen $\vk$-point.
The size of gap $\Delta_{HG}$  in $A^{nonf}(\vk,\omega)$
can be measured by the optical conductivity.
Because $\Delta_{HG}$ has a variation of 70 $\sim$ 100 meV depending on $\vk$-points, 
the measured $\Delta_{HG}$ should show multiplet structures or widespread shape in the optical conductivity measurements.
The small peaks near $E_F$ at 20 K are induced due to the formation of quasiparticle bands 
of $j=5/2$ states within the HG.


\begin{figure}[t]
\begin{center}
\includegraphics[scale=0.22,angle=0]{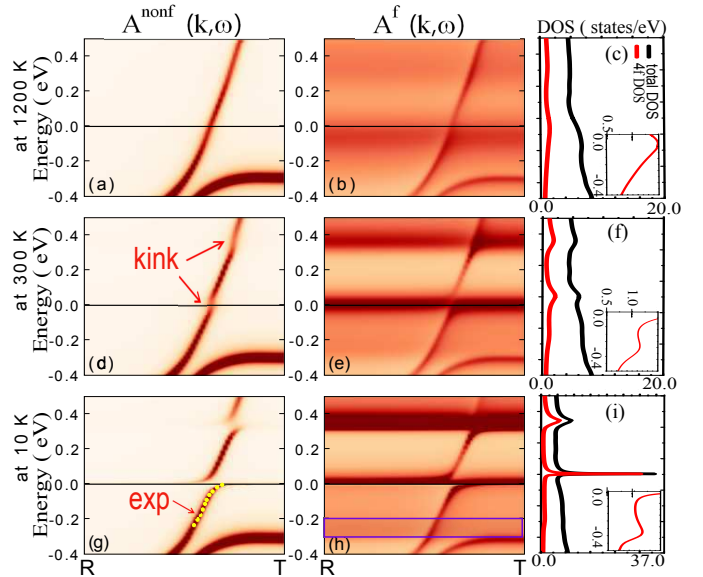}
\end{center}
\caption{(color online)
$T$-dependent $A^{nonf} (\vk,\omega)$ ((a),(d),(g)) and $A^{f} (\vk,\omega)$ ((b),(e),(h)) 
between $R$ $(\pi/2,\pi/2,\pi/2)$ and $T$ $(0,\pi/2,\pi/2)$.
The corresponding DOS's are also given in (c),(f), and (i).
The inset shows the detail of the spin-orbit multiplet at $-300$ meV.
The purple rectangle in (h) indicates the weak spectrum of spin-orbit multiplet,
which can be seen clearly at $-0.3$ eV below $E_{F}$ in (i).
Yellow points in (g) represent the off-resonant ARPES spectra in the measurement,\cite{Im2008}
which are shifted along the $\vk$-path for better alignment of $nonf$ bands.
}
\label{fig2}
\end{figure}

Figure ~\ref{fig2} shows both $A^{nonf}(\vk,\omega)$ 
and the $f$ spectral function ($A^{f}(\vk,\omega)$) along $R$-$T$, and
the integrated density of states (DOS) around $E_{F}$ 
at $T=1200,$ 300, and 10 K.
The $T$-dependent development of the KR states is
clearly confirmed in the DOSs of Figs.~\ref{fig2}(c), (f), (i).
At high $T$, the upper and lower Hubbard bands are located near 2 $\sim$ 3 eV above $E_F$ and 2 eV 
below $E_F$, respectively (not shown here).
At the elevated $T$, the profile of the DOS near $E_F$ comes mostly from $nonf$ states,
although there is a weak background spectrum of $4f$ states.
With lowering $T$, the weights of the upper and lower Hubbard bands are reduced and transferred to the KR states near $E_F$.
The $4f$ states give the main contribution to the DOS near $E_F$ below $T^{*}$.

At high $T$ (1200 K), $A^f (\vk, \omega)$ shows weak intensity near $E_F$,
but has clear dispersive band feature similar to $A^{nonf} (\vk,\omega)$.
This means that small hybridization still exists
between $nonf$ and $f$ states at high $T$.
The broad Gaussian peaks in $A^{nonf} (\vk,\omega)$ at high $T$, as shown in Figs.~\ref{fig4}(c)-(f), are
the indication of this hybridization.
At low $T$ (10 K), where the KR states are fully developed, 
both $A^{f} (\vk,\omega)$ and $A^{nonf} (\vk,\omega)$ show the KR band structures with different weight distribution.
The $nonf$-dominant bands form the gap structure with low intensity at $E_F$, 
while the $f$-dominant bands are confined near $E_F$ to give the sharp KR peak in the integrated DOS,
as shown in Fig.~\ref{fig2}(i).
Note that the KR states should be considered as dispersive band structures,
as described in the periodic Anderson model.

$A^{nonf} (\vk,\omega)$ and $A^{f} (\vk,\omega)$ at low $T$ (10 K) show good agreement 
with the off-resonance and on-resonance ARPES measurements at $T$ = 17 K,\cite{Im2008} respectively.
Especially, the momentum dependence of dispersive KR peaks in the experiments 
is well consistent with the calculated spectrum, as shown in Fig.~\ref{fig2}(g).
Also the weight distribution of $nonf$ and $f$ states observed in experiments are correctly captured in the calculation.
The weak dispersion-less spin-orbit multiplet peak, which exists inside the purple rectangle in Fig.~\ref{fig2}(h),
is also consistent with the experimental observation.
The spin-orbit multiplets of $A^{f}(\vk,\omega)$ are shown around $\pm 0.3$ eV,
and their intensities increase as lowering $T$.
Insets in Figs.~\ref{fig2}(c), (f), (i) provide the $T$-dependent enhancement  of the spin-orbit multiplet around $-$0.3 eV.
It is noteworthy that the multiplet shows almost flat feature 
because the incoherent feature (broadening of bands) is much bigger than the dispersion of the KR states.

%

Because the spectra of high and low $T$ show clearly different quasiparticle band structures near $E_F$, 
the $T$-dependent evolution should show some feature of phase transition or crossover.
Figures~\ref{fig2}(d) and (e) show the spectra in the intermediate $T$. 
$A^{f} (\vk,\omega)$ around $E_F$ shows effectively dispersion-less feature, 
which is the precursor of the formation of the KR states.
Below and above the KR state, the $nonf$ bands are warped in different directions.
At the energy of the KR state, 
the $nonf$ bands are not well defined due to the incoherent contribution of $A^{f} (\vk,\omega)$ to the $nonf$ states.
As a result, $A^{nonf} (\vk,\omega)$ shows the feature of kinks near $E_F$.
Distinctly from the kinks observed in other correlated systems, such as high $T_C$ superconductor,  
the kinks in heavy fermion system should appear in the noncorrelated bands during the formation of the KR bands and the HG.


\begin{figure*}[t]
\includegraphics[scale=0.16,angle=0]{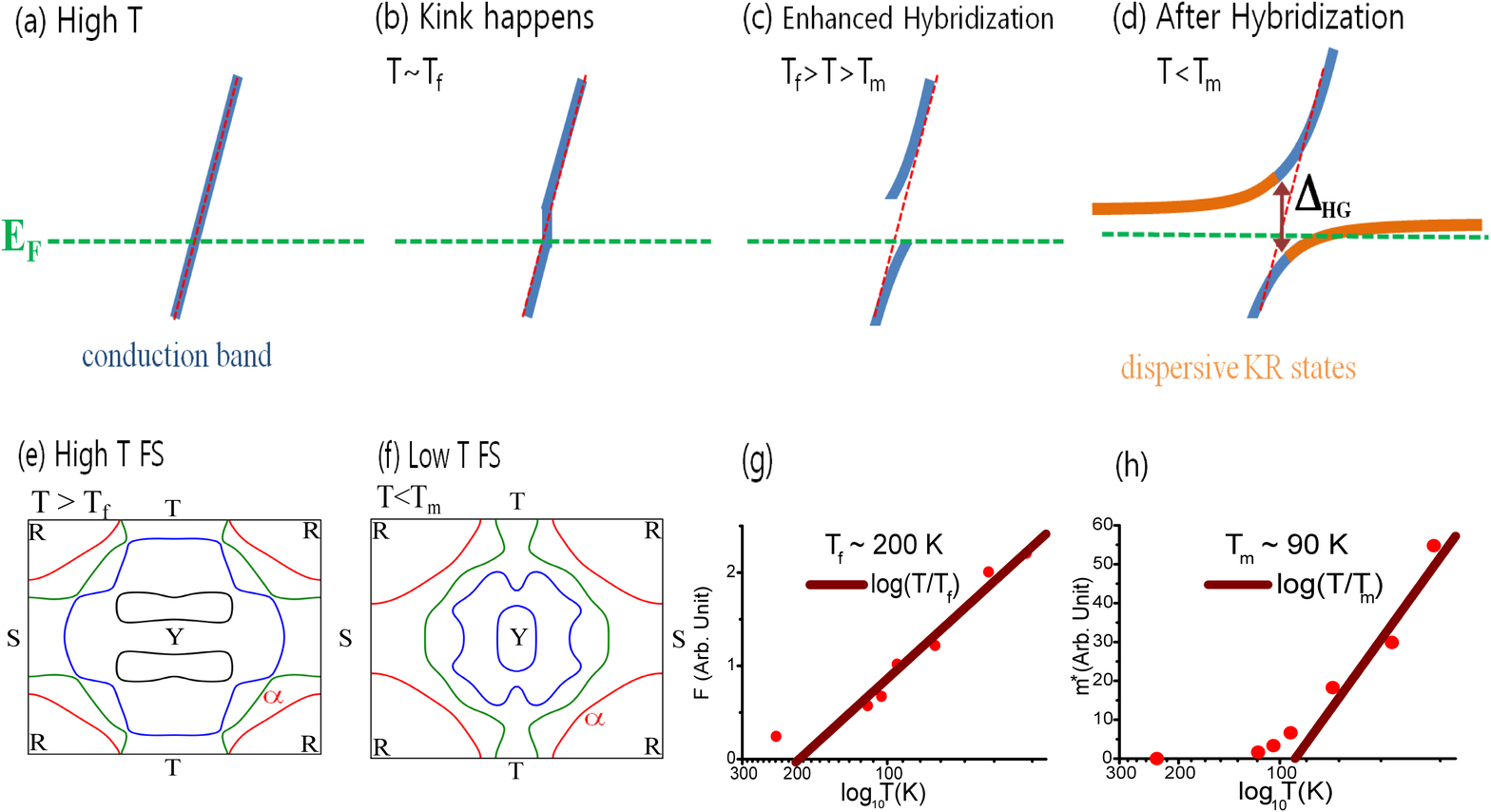}
\caption{(color online)
 The formation of the HG can be divided into four processes in (a)-(d), using  $T_f$ and $T_m$
 that are obtained by the scaling analysis in (g) and (h).
 (a) The high $T$ quasiparticle band is displayed. The blue color represents $A^{nonf} (\vk,\omega)$.
 (b) For $T$ $\sim$ $T_f$, the quasiparticle band is interrupted by the incoherent $4f$ electrons. 
	The red dotted-line is provided as a guide for the high $T$ band.
 (c) For $T_f$ $>$ $T$ $>$ $T_m$, the hybridization between $nonf$ and $4f$ electrons is strengthened.
 (d) For $T$ $<$ $T_m$, the HG and $k$-dependent (dispersive) KR state can be well defined.
 The orange color represents $A^{f} (\vk,\omega)$.
 With lowering $T$, $E_F$ increases gradually due the participation of the new carrier from the incoherent $4f$ electrons.
 The FS on $\vk$=Y plane at high and low $T$ are provided in (e) and (f), respectively. Color represents the different band indices.
 The scaling behaviors of $T$-dependent renormalized de Haas van Alphen frequency (F) and cyclotron effective mass ($m^\ast$) 
	for the FS branch $\alpha$ shown  in (e) and (f) 
 are analyzed in (g) and (h), respectively.
 }
\label{fig5}
\end{figure*}

Figures ~\ref{fig5}(a)-(d) show the schematic picture of emergence of kink during the formation of the KR bands.\cite{supple}
At high $T$ in Fig.~\ref{fig5}(a), 
there are only $nonf$ conduction bands 
that can be usually well described by the open-core band calculation, in which
the occupied Ce $4f$ state is treated as a core level.
With lowering $T$ in Fig.~\ref{fig5}(b),
the incoherent KR states of $4f$ electrons start to contribute to $E_F$,
whereby the kink feature starts to emerge in $A^{nonf} (\vk,\omega)$.
Here the kink structure is far from the "water-fall" shape, rather close to a "bell-profile"" shape,
indicated by the arrow in Fig. 3(d),
since the dispersion changes happen at two spin-orbit multipltes ($j=5/2, 7/2$).
As decreasing $T$ further, the $f$ electrons start to be coherent slowly, and the $nonf$ bands are still being deformed.
This process corresponds to Fig.~\ref{fig5}(c), where the coherent character of bands becomes enhanced around $E_F$.
In this case, $A^{nonf} (\vk,\omega)$ has the kink structure of the "water-fall" shape 
due to the separation of the upper and lower hybridized bands.
At lower $T$ in Fig.~\ref{fig5}(d),
most $4f$ electrons near $E_F$ become coherent to make the fully coherent bands near $E_F$.
Accordingly, the region of the kink feature is changed into that of the HG feature.
Interestingly, the electron FS area gradually enlarges during this procedure.

The area of the electron FS around $R$, 
which is identified as $\alpha$ branch in Figs.~\ref{fig5}(e) and (f), 
increases continuously with lowering $T$.
In our recent DMFT study\cite{hcc} on the FS of heavy fermion CeIrIn$_5$,
two temperature scales are proposed: one ($T_f$) for the $T$-dependent evolution of the FS size,
and the other ($T_m$) for the $T$-dependent evolution of the cyclotron effective mass ($m^\ast$).
$T_f$ should be related to the contribution of local $4f$ electron to conduction electron.
On the other hand, $T_m$ is a characteristic of the formation of coherent $4f$ bands
in the lattice since the $m^\ast$ reflects the renormalization of the carriers.
Although $T_m$ is defined by the change of the effective mass of the FS, 
it should be similar to $T^*$ where the the formation of the coherent KR states begins.\cite{hcc} 
Similar to CeIrIn$_5$, the same scaling behavior 
is also shown in CeCoGe$_2$.
The FS branch $\alpha$, which is  the well-defined FS branch at all $T$, as shown in Figs. \ref{fig5}(e) and (f),
is  chosen for this study.
By analyzing $T$-dependent scaling behaviors in Figs.~\ref{fig5}(g) and (h),
we found $T_f \sim $ 200 K and $T_m \sim$ 90 K, respectively, for $\alpha$ branch. 

The kink can be observed around $T_f$, where the incoherent $f$ state contributes to $E_F$. (see the Supplementary Movie.)
The kink phenomenon is changed into the gap feature
gradually between $T_f$ and $T_m$,
where the contribution of incoherent $4f$ electron states disturb the band dispersion near $E_F$.
Well below $T_m$, the HG and KR states are well defined.
So, the formations of the kink around $T_f$ (200 K) will be the precursor of the HG below $T_m$ (90 K).
Note that the kink features are also observed around 0.3 eV above $E_F$,
as shown in Fig 3.(d),
due to the incoherent contribution of spin-orbit multiplet. 
The multiplet around $-$0.3 eV does not give the kink because
the contribution of $f$ state is too weak to distort the $nonf$ bands.


In summary, we have analyzed the $T$-dependent evolutions of $A(\vk,\omega)$ in the heavy fermion compound CeCoGe$_2$.
It is shown that the DFT+DMFT calculations are consistent with the experimental measurements.
We propose that
the kink of $A^{nonf} (\vk,\omega)$ around $E_F$ can be identified during
the evolution from the dispersive $nonf$ state at high $T$ to the HG and KR states at low $T$.
Phenomenologically, the kinks observed in this work will show the similar shape
to those in other experiments, even though
the conventional kinks appear in the correlated bands
via the interaction with other excitations, such as phonon and spin-fluctuation.
The kinks in current work should be distinguished also 
from the one only with the electronic correlation in previous studies.\cite{Byczuk,Kainz,Held}
As indicated in Fig. 4, 
the kink can be observed between $T_m$ and $T_f$, 
while all other kinks in previous studies should be shown well below $T^{*}$ ($\sim T_m$ ).
The kink induced by the correlation
between the incoherent $4f$ and dispersive $nonf$ electrons above $T^{*}$
can be investigated in the state-of-the-art $T$-dependent ARPES experiments.
We suggest that
the detailed analysis on the abrupt change of electron velocity and the scattering rate near the
Fermi level will provide crucial information for
the heavy fermion system.


\begin{acknowledgments}
We acknowledge useful discussions with Hojun Im (Hirosaki University) and Tuson Park  (Sungkyunkwan University).
This work was supported by the National Research Foundation of Korea(NRF) funded by the Ministry of Education, Science and Technology 
(No. 2009-007994, 2010-0006484, 2010-0026762, 2012029709).
\end{acknowledgments}

\end{document}